\documentclass[aps,prl,twocolumn,10pt,showpacs,showkeys]{revtex4-2}
\usepackage[utf8]{inputenc}
\usepackage{amsmath}
\usepackage{bm}
\usepackage{natbib}
\usepackage{graphicx}
\usepackage[separate-uncertainty=true]{siunitx}
\usepackage[nolist]{acronym}
\usepackage{color}

\begin{document}
\DeclareSIUnit\angstrom{\text {Å}}
\DeclareSIUnit\bar{bar}
\title{Host-atom-driven transformation of a honeycomb oxide into a dodecagonal quasicrystal}
\author{Martin Haller}
\author{Julia Hewelt}
\author{V. Y. M. Rajesh Chirala}
\author{Loi Vinh Tran}
\author{Ankur Bhide}
\author{Muriel Wegner} 
\author{Stefan F\"orster}
\email{stefan.foerster@physik.uni-halle.de}
\author{Wolf Widdra}
\affiliation{Institute of Physics, Martin-Luther-Universit\"at
Halle-Wittenberg, D-06099 Halle, Germany}

\date{\today}

\begin{abstract} 

Dodecagonal oxide quasicrystals (OQCs) have so far been limited to a few elemental systems, with no general formation mechanism established. Here, we demonstrate a versatile approach to OQC formation via a host-atom-induced transformation of a metal-oxide honeycomb (HC) network. Adsorption of Ba, Sr, or Eu onto the HC layer triggers its reorganization into a dodecagonal tiling, as revealed by low-energy electron diffraction and scanning tunneling microscopy. Full conversion occurs when \SI{73}{\percent} of the honeycomb rings are occupied. Kelvin probe and UV photoelectron spectroscopy show a linear decrease in work function with increasing host coverage, followed by a sharp increase upon quasicrystal formation due to reduced host dipoles. This transformation mechanism enables the fabrication of structurally precise OQCs, including a new Eu-Ti-O phase that extends the field to lanthanide quasicrystals, forming a 2D grid of localized magnetic moments. The method offers a general route to explore lattice-matched substrates for epitaxial growth and may be adapted to other 2D honeycomb materials—such as graphene, hexagonal ice, and silica—paving the way for engineered aperiodic systems beyond transition metal oxides.

\end{abstract}


\maketitle  

\begin{figure}[t]%
	\includegraphics[width=\linewidth]{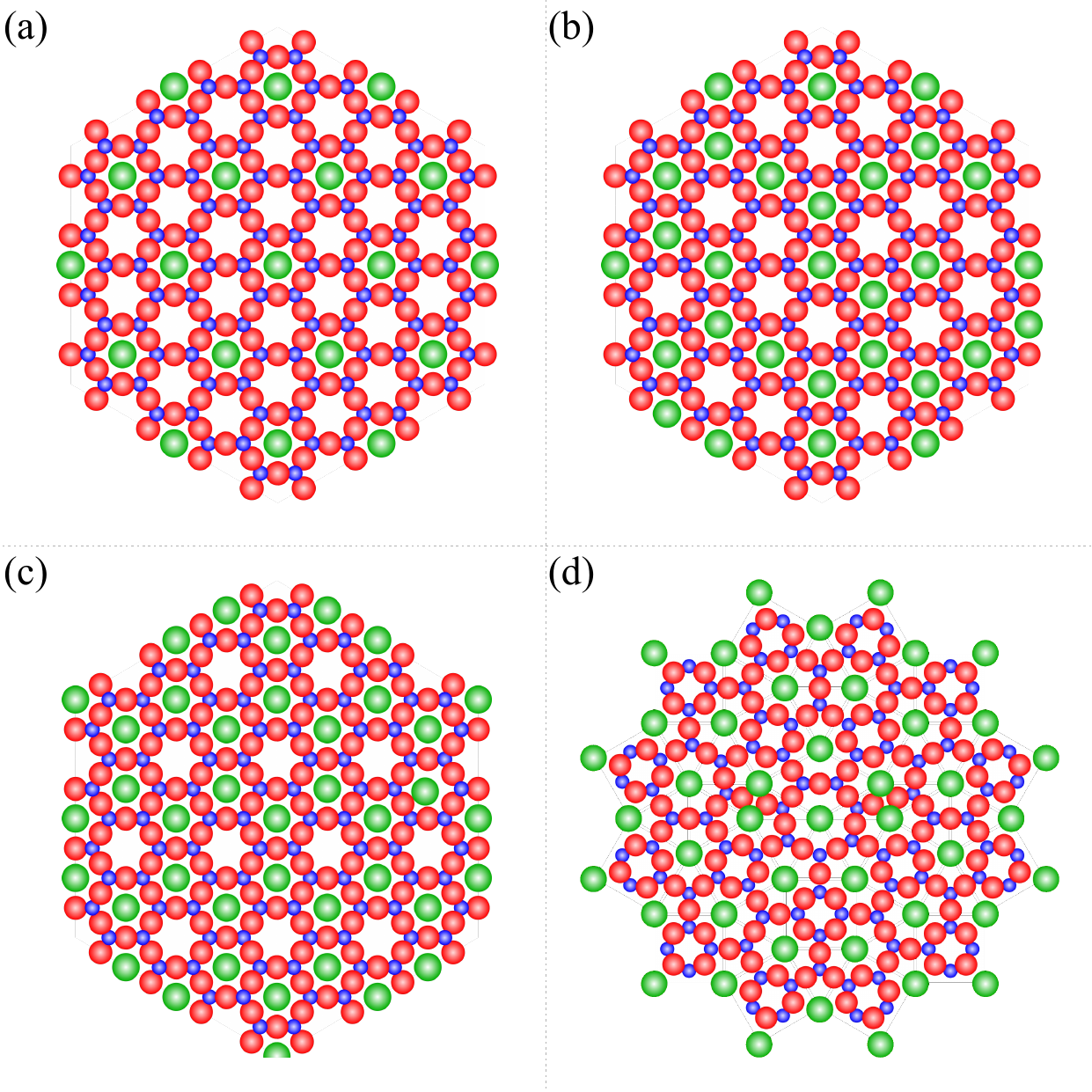}
	\caption{(a-c) Schematic model of a metal-oxide honeycomb (HC) structure (blue and red circles) with increasing atom decoration (green circles) with a HC ring coverages of \SIlist{33;50;67}{\percent}. The host atoms considered here are Ba, Sr, or Eu. The atoms are set to scale in proportion to the atomic radii of O, Ti, and Ba for six-fold coordination \cite{Mueller_anorganische_2008}. A ($\sqrt3\times\sqrt3)\textit{R30°}$ superstructure arises from (a) the occupied and (c) the empty HC pores. (b) Labyrinth phase formed at \SI{50}{\percent}. (d) Schematic model of the dodecagonal structure in which the host atoms decorate the vertices of a square-triangle-rhombus tiling. These host atoms are embedded in a Ti$_n$O$_n$ rings network with n=4, 7, and 10, in which \SI{73}{\percent} of the rings are occupied.}
	\label{fig:Models}
\end{figure}

Quasicrystals (QCs) are unique materials with long-range order and a discrete diffraction pattern, but without periodicity or repeating unit cells as found in conventional periodic crystals. First identified in intermetallic alloys in 1984, QCs exhibit unconventional symmetries such as 5-, 8-, 10-, or 12-fold rotational symmetry \cite{shechtman1984}. Since then, QCs have been discovered in a variety of materials including liquid crystals \cite{zeng2023}, mesoporous silica \cite{xiao2012}, and even macroscopic hard spheres \cite{plati2024}. Their non-periodic structure gives rise to intriguing magnetic properties, such as enhanced antiferromagnetic ordering \cite{jagannathan2012a} and unique spin-wave propagation mediated by phason excitations, a characteristic mode of QCs \cite{szallas2009,deboissieu2012}. These properties have attracted increasing attention following the recent discovery of ferromagnetic ordering in an intermetallic QC \cite{tamura2021}. 
A major breakthrough in the field of quasicrystals occurred a decade ago when the first 2D materials with aperiodic atomic order, namely two-dimensional oxide quasicrystals (OQCs) featuring dodecagonal symmetry, were discovered \cite{forster2013}. 
Recent findings reveal a strong connection between OQCs and 2D oxide honeycomb (HC) structures \cite{schenk2022}. The atomic framework of OQCs consists of trivalent metals in metal - oxygen rings,  M$_n$O$_n$, with ring sizes n = 4, 7, and 10 where rings of n = 7 and 10 are stabilized by additional (alkaline earth) host atoms as depicted in Fig. \ref{fig:Models}(d) \cite{cockayne2016,schenk2022}. The rings form a square-triangle-rhombus tiling, which follows the dodecagonal quasicrystal model of Niizeki and Gähler \cite{niizeki1987,gahler1988,schenk2019}. The structure can be derived from the HC lattice through Stone-Wales transformations \cite{stone1986, schenk2022}, which create a network of 4- and 7-membered rings, with 10-membered rings emerging from pore expansion via oxygen incorporation.
 
Earlier research on the decoration of titanium oxide honeycombs with Ba demonstrated that Ba host atoms occupy the rings of the HC structure and organise themselves to maximize separation from neighboring Ba atoms \cite{wu2015}. Density functional theory (DFT) calculations indicate that these host atoms transfer their valence electrons to the oxide layer and reside within the pores as positively-charged ions \cite{dorini2021,wu2015}.
Thus, it is the repulsive interaction between neighboring host atoms that drives self-organization. As a consequence, at a coverage of one third of all pores, a $(\sqrt{3}\times\sqrt{3})\textit{R30°}$ superstructure forms. At \SI{50}{\percent} coverage, a labyrinth phase is observed. At a coverage of two thirds, again a $(\sqrt{3}\times\sqrt{3})\textit{R30°}$ superstructure forms, which fills remaining free HC pores and hence reducing the next-neighbor distance from $\sqrt{3} a_{HC}$ to $a_{HC}$ (Fig. \ref{fig:Models}(a-c)). Monte-Carlo simulations involving first-, second-, and third-neighbor effective Ba-Ba interactions perfectly reproduce the experimental findings. From the successive increase in the density of host atoms at the surface, a linear decrease of the workfunction of the systems was predicted \cite{wu2015}. 

The dodecagonal oxide quasicrystal network is characterized by a coverage of \SI{73}{\percent} of all available Ti$_n$O$_n$ rings with host atoms. The benefit of converting a honeycomb network to a dodecagonal OQC network is a \SI{15}{\percent} increase in the next-neighbor distance between host atoms residing in the n=7 rings in comparison to $a_{HC}$ (Fig. \ref{fig:Models}(d)). Hence, the next-neighbor repulsion is reduced \cite{schenk2022}. Furthermore, the larger ring sizes allow a reduced height of the host atoms above the substrate and thus a reduced strength of their surface dipoles \cite{schenk2022}. 

In this letter, we present a generalized approach for the formation of dodecagonal metal-oxide quasicrystals by the transformation of a metal-oxide HC via decoration with host metal atoms. It is demonstrated here for Eu as the host atom species, but can be extended to other host atoms as well. Our finding paves the way for a large variety of dodecagonal metal-oxide quasicrystals, such as those from the family of lanthanides with interesting magnetic properties in frustrated and aperiodic structures.

\begin{figure}[t]
	\includegraphics[width=\linewidth]{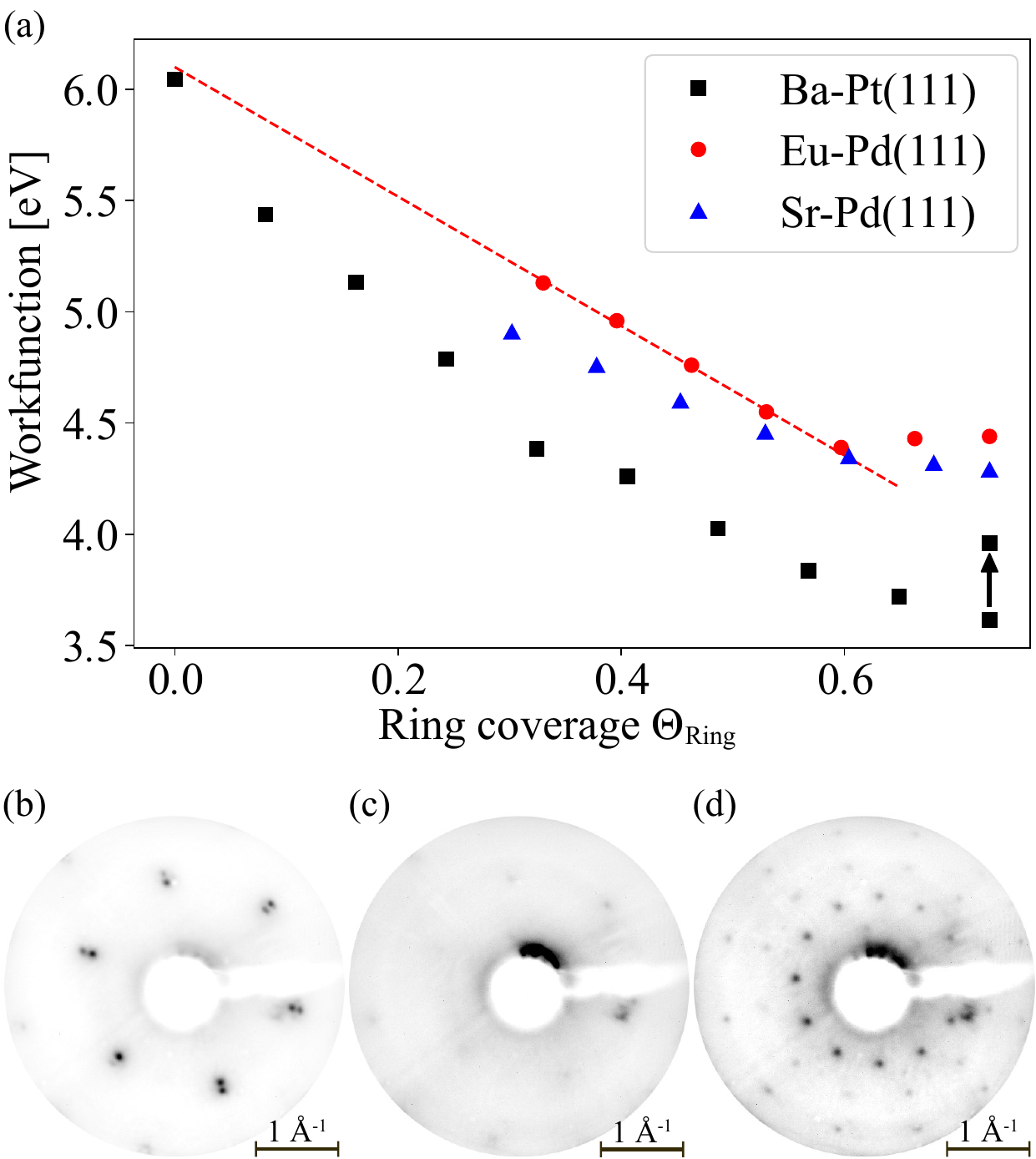}
	\caption{(a) The decrease in the ring coverage-dependent workfunction of a titanium oxide HC layer grown on Pt(111) and Pd(111) substrates upon deposition of Ba (black squares), Eu (red circles) and Sr(blue triangles). The dashed line depicts the linear model describing the workfunction decrease for the Eu-Pd(111) series. The black arrow indicates the workfunction increase induced by the transformation of the Ba-decorated HC to the OQC network. Diffraction patterns of (b) the pristine HC, (c) the Ba-decorated HC before and (d) after conversion to the OQC.}
	\label{fig:Workfunction}
\end{figure} 

\begin{figure*}[t]
	\includegraphics[width=\linewidth]{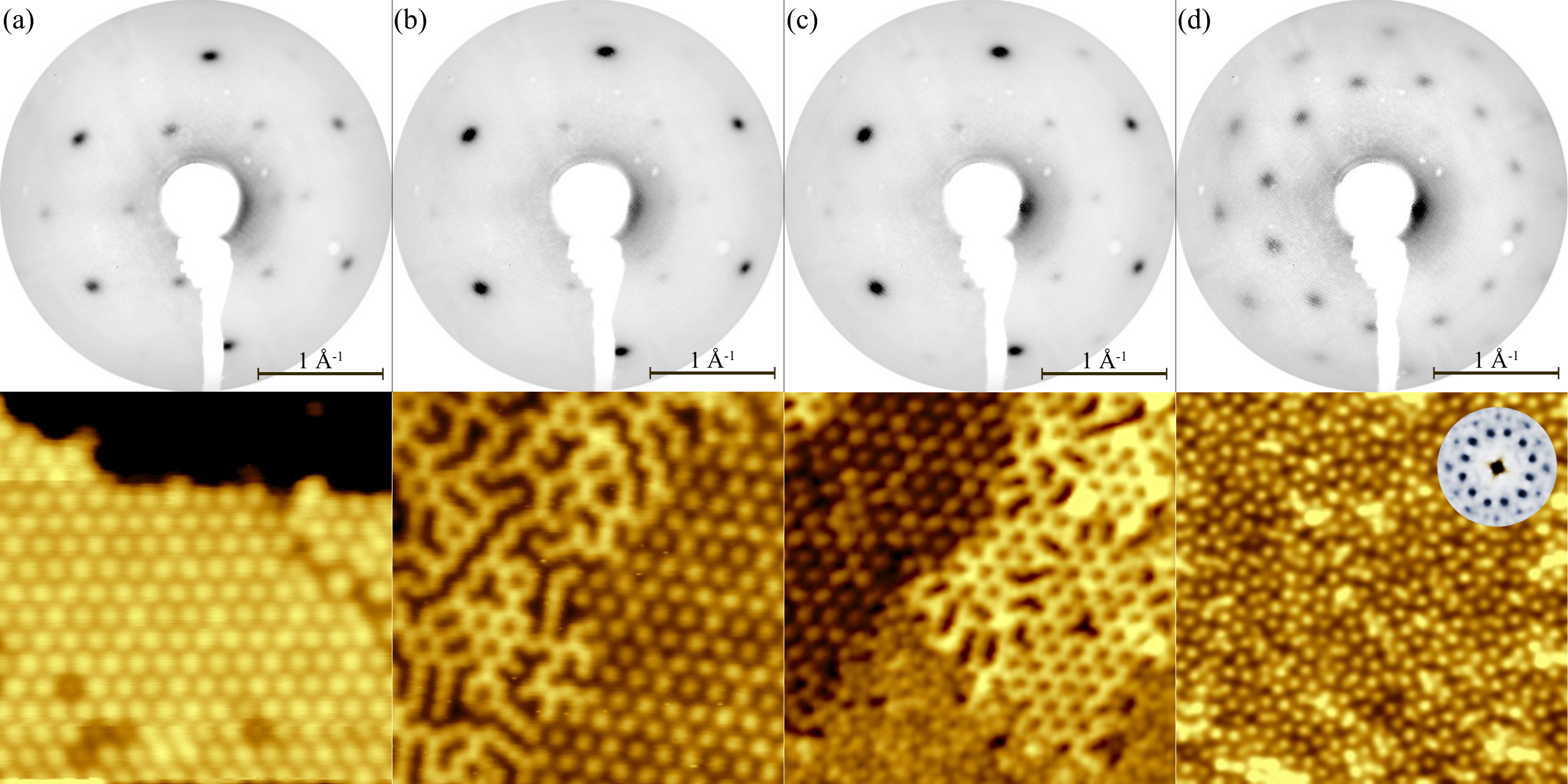}
	\caption{LEED (top) and STM data (bottom) of an Eu-covered Ti-O honeycomb layer on Pd(111) at ring coverages of (a) \SI{33}{\percent}, (b) \SI{46}{\percent}, (c) \SI{53}{\percent} and (d) \SI{73}{\percent}. The six intense outer spots in (a-c) correspond to the HC structure. The additional spots originate from a ($\sqrt3\times\sqrt3$)\textit{R30°} superstructure at \SIlist{33;67}{\percent} coverage (Fig. \ref{fig:Models}(a,c)). (d) At \SI{73}{\percent} ring coverage, the network transforms to a dodecagonal square-triangle-rhombus tiling. The inset shows the FT of atom positions extracted from the STM image. STM images are 15$\times$15 nm² each. (a) \SI{1}{\volt}, \SI{1}{\nano\ampere}. (b) \SI{-2}{\volt}, \SI{0.1}{\nano\ampere}. (c) \SI{1}{\volt}, \SI{0.3}{\nano\ampere}. (d) \SI{0.4}{\volt}, \SI{1}{\nano\ampere}.}
	\label{fig:EuPdStructure}
\end{figure*} 

The decoration and conversion of a titanium oxide HC structure upon step-wise deposition of host atoms with annealing steps in between is characterized by monitoring the workfunction changes using area integrating UV photoelectron spectroscopy (UPS) or Kelvin-probe measurements (KP). Annealing is performed to promote the adatom mobility \cite{wu2015}. Ba, Sr and Eu are supplied by thermal evaporation of the metallic compound from a home-built evaporation source. On Pt(111), titanium oxide films are grown by molecular beam epitaxy of \SI{1.5}{\angstrom} of Ti with subsequent oxidation to Ti$^{4+}$ at \SI{E-5}{\milli\bar} and \SI{900}{\kelvin} followed by \SI{10}{\min} annealing at \SI{1050}{\kelvin} in ultrahigh vacuum (UHV). As a result, \SI{90}{\percent} of the Pt(111) surface is covered with the HC structure forming a $(2.15\times2.15)$ superstructure, which corresponds to a HC lattice parameter of \SI{5.95}{\angstrom} as reported earlier \cite{sedona2005}. Onto this layer, Ba is deposited with a subsequent annealing for \SI{5}{\min} at \SI{670}{\kelvin} in UHV.
On Pd(111), the pristine HC structure is intrinsically unstable and converts to either a more reduced ''zigzag'' or the higher oxidized ''rectangular'' structure \cite{farstad2016}. However, we find that subsequent deposition of Sr or Eu with post-annealing at \SI{870}{\kelvin} in \SI{5E-10}{\milli\bar} O$_2$ stabilizes the titanium oxide HC structure on Pd(111) for a HC ring coverage of \SI{33}{\percent}. The unit cell structure is rotated by \SI{30}{\degree} against the Pd(111) substrate and possesses a slightly-reduced lattice spacing of \SI{5.80}{\angstrom} as compared to Pt(111). 

Figure \ref{fig:Workfunction}(a) shows the workfunction changes of a titanium oxide HC upon decoration with Ba, Sr, and Eu. The pristine HC layer posseses a workfunction of \SI{6.05\pm0.05}{\electronvolt} when grown on Pt(111) (black squares). In diffraction, this HC shows a clear six-fold pattern (Fig. \ref{fig:Workfunction}(b)). 
Upon decorating the HC with Ba, a rapid drop in the workfunction is observed. This drop is almost linear up to a coverage of $\sim$ \SI{30}{\percent} of the HC rings. For higher coverages it continues with a slightly reduced linear slope. The conversion of the decorated HC into the OQC is induced by annealing the film at \SI{1070}{\kelvin} in the presence of \SI{5E-9}{\milli\bar} of O$_2$. Upon transformation, the twelvefold pattern of sharp diffraction spots prove the formation of the dodecagonal quasicrystal (Fig. \ref{fig:Workfunction}(d)). This structural transformation is accompanied by an increase in workfunction of \SI{0.35}{\electronvolt} to a final value of \SI{3.95\pm0.05}{\electronvolt} (black arrow in Fig. \ref{fig:Workfunction}(a)).

The same approach can be applied for the formation of an Eu-based OQC. Since the ionic radius of Eu is smaller as compared to Ba, experiments carried out on Pd(111) provided a substrate lattice spacing that is reduced by \SI{0.8}{\percent} compared to Pt(111). By depositing the equivalent of $\sim$ \SI{33}{\percent} ring coverage to the pre-deposited titanium oxide layer and annealing at \SI{870}{\kelvin} in UHV, the hexagonal structure of the titanium oxide HC is observed in LEED. In addition, a $(\sqrt{3}\times\sqrt{3})$\textit{R30°} superstructure is visible that results from Eu decorating one third of the HC pores as depicted in Fig. \ref{fig:Models}(a). The corresponding LEED pattern along with the local atomic arrangement determined by STM is shown in Fig. \ref{fig:EuPdStructure}(a). 
The workfunction of the layer amounts to \SI{5.1}{\electronvolt} (red circles in Fig. \ref{fig:Workfunction}(a)), which is \SI{0.7}{\electronvolt} higher in comparison to the Ba-decorated titanium oxide HC. 
By increasing the Eu coverage in increments of \SI{7}{\percent}, the workfunction drops linearly for coverages between \SIlist{33;60}{\percent} before it saturates. 
The linear decrease of the workfunction can be rationalized by using the formula $\Phi(\Theta_{Ring})=\Phi_{HC}-\frac{e}{\epsilon_0 A_{uc}}p_z\Theta_{Ring}$, where e denotes the elementary charge, $\epsilon_0$ is the electrical field constant, and $A_{uc}$ is the area of the honeycomb unit cell. The dipole moment perpendicular to the surface of one host atom $p_z$ is assumed to be a constant. By extrapolating to the absence of host atoms, we derive $\Phi_{HC}=6.10\pm0.04$ eV that is in good agreement with the experimental value obtained on Pt(111) (red dashed line in Fig. \ref{fig:Workfunction}(a)). From the slope of the line of \SI{-2.91\pm0.09}{\electronvolt}, a dipole moment per Eu ion of 2.36 Debye is determined.
Figure \ref{fig:EuPdStructure}(b) shows that above a coverage of $\Theta_{Ring}=1/3$, the $(\sqrt{3}\times\sqrt{3})$\textit{R30°} superstructure spots lose intensity in LEED due to an increasing area fraction covered by the labyrinth phase that represents a ring coverage of \SI{50}{\percent}. The corresponding STM image confirms this phase co-existence. However, for coverages exceeding \SI{50}{\percent} Fig. \ref{fig:EuPdStructure}(c) shows that the labyrinth phase is strongly suppressed and two different $(\sqrt{3}\times\sqrt{3})$\textit{R30°} superstructures are found in co-existence, corresponding to 1/3 and 2/3 ring coverage. Consequently, the superstructure spots gain in intensity again. 
In parallel to this structural evolution, the transformation of the hexagonal HC rings proceeds at a reduced speed. At a coverage of \SI{53}{\percent}, we notice that \SI{6}{\percent} of the surface area already exhibit a square-triangle-rhombus tiling (bottom of Fig. \ref{fig:EuPdStructure}(c)). At a nominal decoration of \SI{60}{\percent}, the square-triangle-rhombus tiling covers already \SI{50}{\percent} of the 2D oxide layer. For even higher coverage, the OQC is the dominant phase in LEED as shown for a coverage of \SI{73}{\percent} in Fig. \ref{fig:EuPdStructure}(d). The continuous progression of the network transformation over multiple deposition steps is the reason for the constant workfunction of the system above \SI{60}{\percent} coverage: The decreasing workfunction as the Eu density rises steadily is balanced by the increase in workfunction resulting from the transformation of the Ti-O network. An estimation of the workfunction difference between the decorated honeycomb at \SI{73}{\percent} coverage and the OQC is given by the difference of the measurement and the linear fit of the data. At a coverage of \SI{73}{\percent} the fit predicts a workfunction of \SI{3.98\pm0.08}{\electronvolt}, whereas our experimental value for the OQC is \SI{4.44\pm0.06}{\electronvolt}. Hence, the difference amounts to \SI{0.46\pm0.10}{\electronvolt}, which is comparable to the Ba-Pt(111) case. 

\begin{figure}[t]
	\includegraphics[width=\linewidth]{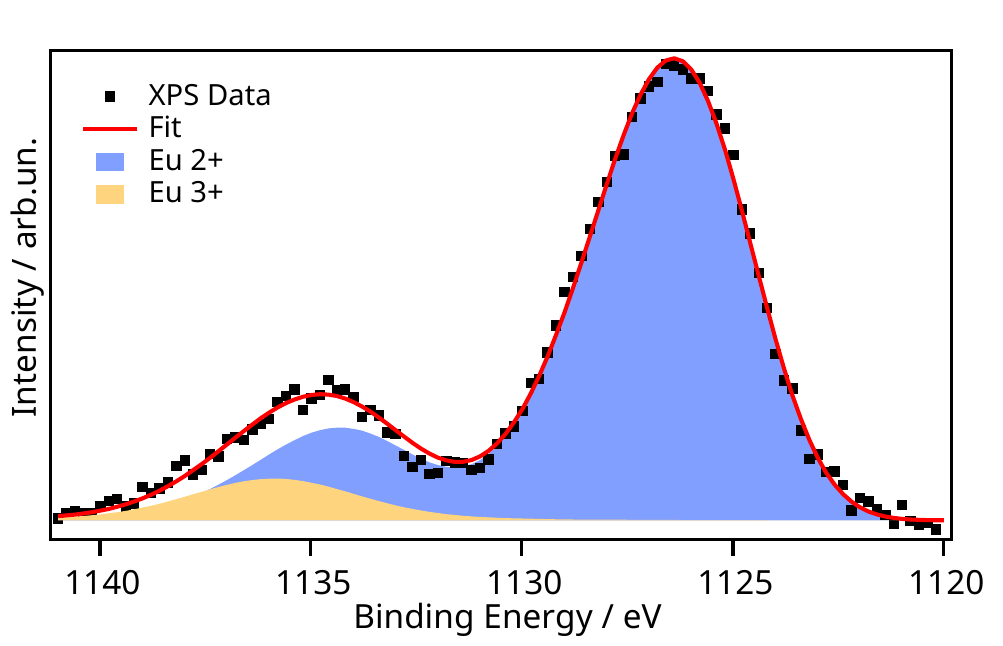}
	\caption{Eu\textit{3d$_{5/2}$} XPS data recorded for the OQC in Eu-Ti-O/Pd(111). The predominant oxidation state of Eu is +2. Hence Eu resides in a high-spin state in the aperiodic network.}
	\label{fig:EuXPS}
\end{figure}
In contrast to the alkaline earth metals, Eu is able to exhibit multiple oxidation states: In the +2 valence state the 4\textit{f} shell is half-filled, which corresponds to a high-spin configuration. Alternatively, the oxidation to a +3 valence state is possible, corresponding to a low-spin configuration of the Eu atoms. The core-level photoemission data of the Eu-derived OQC on Pd(111) shown in Figure \ref{fig:EuXPS} unambiguously proofs that the Eu host atoms adopt a +2 valence in the OQC, thus behaving chemically similar to the alkaline earth metals. These data suggest that the Eu-derived OQC represents an aperiodically-ordered ensemble of large magnetic moments.  

\begin{figure}[t]
	\includegraphics[width=\linewidth]{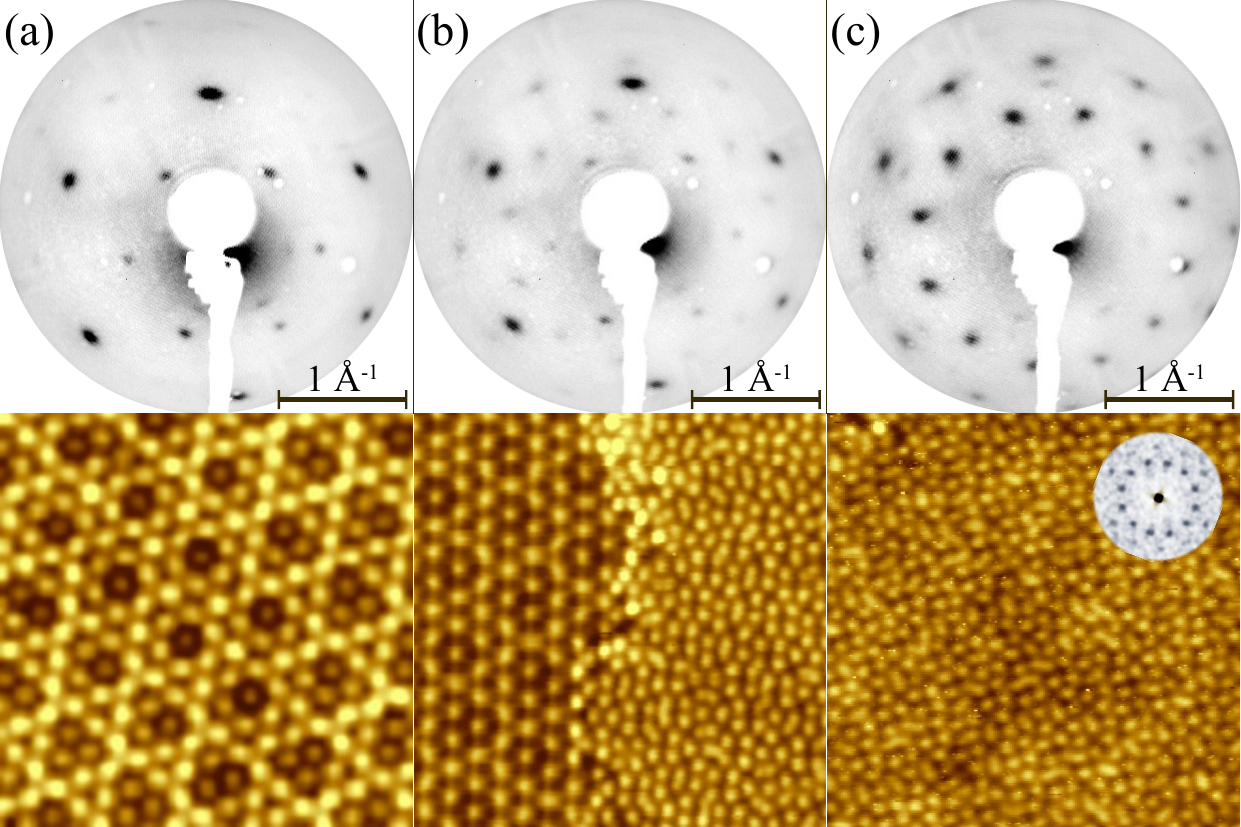}
	\caption{LEED (top) and STM data (bottom) for a Sr covered Ti$_2$O$_3$ honeycomb layer on Pd(111) at (a) \SI{33}{\percent}, (b) \SI{46}{\percent} and (c) \SI{73}{\percent} coverage. 
	The six intense outer spots at the lowest coverage originate from the HC layer. The additional ($\sqrt3\times\sqrt3)\textit{R30°}$ superstructure spots relate to the Sr ions decorating 1/3 of the HC pores. (b) For higher coverage, a coexistence of the ($\sqrt3\times\sqrt3)\textit{R30°}$ decorated HC structure and the OQC is observed. (c) At 0.73 ring coverage, the HC almost completely disappears due to the conversion to the dodecagonal oxide quasicrystal. The inset shows the FT of the atom positions extracted from the STM image. 
	STM images 15$\times$15 nm² each. (a) \SI{1.5}{\volt}, \SI{0.3}{\nano\ampere}. (b) \SI{-0.4}{\volt}, \SI{0.3}{\nano\ampere}. (c) \SI{-0.4}{\volt}, \SI{0.7}{\nano\ampere}.}
	\label{fig:SrPdStructure}
\end{figure}

So far we have discussed the generalized approach for dodecagonal metal-oxide quasicrystals by the transformation of a metal-oxide HC due to the decoration with Eu as host atoms. In the following, we use this approach for Sr as host atoms.
The OQC in this system was found to develop as a minority species when grown on Pt(111) with a competing \textit{periodic} square-triangle-rhombus structure that grows commensurately and epitaxially stabilized on Pt(111) \cite{schenk2017}. 
Motivated by a high-structural quality of the OQC in Eu-Ti-O/Pd(111) and considering almost identical ionic radii of Sr and Eu, we pursued the decoration of a HC with Sr on the Pd(111) substrate. The change in the workfunction plotted against the Sr coverage (blue triangles in Fig. \ref{fig:Workfunction}(a)) indicates strong similarities to the characteristics found for Eu. There is again an almost linear drop in the workfunction of the Sr-Ti-O/Pd(111) system up to a coverage of \SI{60}{\percent} at a slightly-reduced slope compared to Eu. By further increasing the Sr coverage, the workfunction shows only a minor reduction to a final value of \SI{4.28\pm0.06}{\electronvolt}. Figure \ref{fig:SrPdStructure} shows that in the Sr-Ti-O/Pd(111) system, the decoration of the HC also results in the formation of a ($\sqrt3\times\sqrt3)$\textit{R30°} superstructure at a coverage of \SI{33}{\percent}. On the atomic scale, the ($\sqrt3\times\sqrt3)\textit{R30°}$ exhibits a complex height modulation of the host atoms, which has been seen before for encapsulated Pd(111) nano islands grown on SrTiO$_3$ \cite{silly2005}. 
In contrast to Eu, we do not observe the formation of a labyrinth phase or the ($\sqrt3\times\sqrt3)\textit{R30°}$ superstructure related to a ring coverage of 2/3 upon increasing the Sr coverage. Instead, LEED and STM reveal an immediate transformation of the HC to a well-ordered dodecagonal OQC when the coverage exceeds \SI{33}{\percent} (Fig. \ref{fig:SrPdStructure}(b)). When the ring coverage approaches \SI{73}{\percent}, an almost complete transformation to the OQC is observed. In the corresponding LEED pattern in Fig. \ref{fig:SrPdStructure}(c), the HC can still be recognized. However, the diffraction pattern is dominated by the sharp spots of the OQC structure. On the atomic scale, the dodecagonal structure is seen from the Fourier Transform of the atom positions in the STM image, which is shown as an inset in Fig. \ref{fig:SrPdStructure}(c). This series of experiments directly proves that a well-ordered OQC can form in Sr-Ti-O when the appropriate substrate is chosen. 

In conclusion, the experiments presented here prove that oxide quasicrystals of excellent structural quality can be fabricated by a network transformation of a binary oxide HC decorated with a host atom species. Following this approach a new OQC phase is discovered upon Eu deposition on titanium oxide, which broadens the field of two-dimensional aperiodic materials towards lanthanides. Thus, a two-dimensional dodecagonal grid of localized magnetic moments is fabricated. Furthermore, this well-controlled network transformation approach is  well-suited for the exploration of optimal substrate lattice matching conditions for epitaxial growth of long-range ordered OQC phases as demonstrated for the Sr-Ti-O system. This adatom-assisted network-transformation mechanism might be adapted to drive 2D honeycomb systems beyond Ti-O into quasicrystals, especially HC systems which are known to exhibit intrinsic defects associated with the Stone-Wales transformation. Therefore, we speculate that two-dimensional structures such as graphene \cite{robertson2012,kurasch2012}, hexagonal ice \cite{ma2020,hong2024}, silicene \cite{sahin2013}, silica \cite{zhong2022} or iron-vanadium oxides \cite{wemhoff2022} might be  driven into ring geometries that support 2D quasicrystals.

We thank R. Kulla for his technical support. This work was funded by the European Union (EFRE) and the Deutsche Forschungsgemeinschaft (DFG, German Research Foundation) – 406658237.

\bibliography{HC-to-OQC-Paper.bib}

\end{document}